\documentclass[aps,prc,twocolumn,preprintnumbers,amsmath,amssymb,superscriptaddress,nofootinbib]{revtex4-2} 

\usepackage{amsthm}
\usepackage{amssymb}
\usepackage{mathtools}
\usepackage{physics}
\usepackage{graphicx}
\usepackage{adjustbox}
\usepackage{tabularx}
\usepackage{placeins}
\usepackage{booktabs}
\usepackage{multirow}
\usepackage{graphicx}
\usepackage{dcolumn}
\usepackage{bm}
\usepackage{color}
\usepackage[dvipsnames]{xcolor}
\usepackage{comment}
\usepackage[pdftex,colorlinks,citecolor=blue,urlcolor=blue,linkcolor=blue,bookmarks]{hyperref}

\begin{document}
\title{Neutrinoless $\beta\beta$-decay nuclear matrix elements from two-neutrino $\beta\beta$-decay data}

\author{Lotta Jokiniemi}
\email{ljokiniemi@triumf.ca}
\affiliation{Departament de Física Quàntica i Astrofísica, Universitat de Barcelona, 08028 Barcelona, Spain}
\affiliation{Institut de Ciències del Cosmos, Universitat de Barcelona, 08028 Barcelona, Spain}
\affiliation{TRIUMF, 4004 Wesbrook Mall, Vancouver, BC V6T 2A3, Canada}
\author{Beatriz Romeo}
\email{bromeo@dipc.org}
\affiliation{Donostia International Physics Center, 20018 San Sebasti\'an, Spain}
\author{Pablo Soriano}
\email{pablosoriano@ub.edu}
\affiliation{Departament de Física Quàntica i Astrofísica, Universitat de Barcelona, 08028 Barcelona, Spain}
\affiliation{Institut de Ciències del Cosmos, Universitat de Barcelona, 08028 Barcelona, Spain}
\author{Javier Men\'{e}ndez}
    \email{menendez@fqa.ub.edu}    
\affiliation{Departament de Física Quàntica i Astrofísica, Universitat de Barcelona, 08028 Barcelona, Spain}
\affiliation{Institut de Ciències del Cosmos, Universitat de Barcelona, 08028 Barcelona, Spain}

\date{\today}

\begin{abstract}
We study two-neutrino ($2\nu\beta\beta$) and neutrinoless double-$\beta$ ($0\nu\beta\beta$) decays in the nuclear shell model and proton-neutron quasiparticle random-phase approximation (pnQRPA) frameworks. Calculating
the decay half-life of several dozens of nuclei ranging from calcium to xenon with the shell model,
and of $\beta\beta$ emitters with a wide range of proton-neutron pairing strengths in the pnQRPA, we observe good linear correlations between $2\nu\beta\beta$- and $0\nu\beta\beta$-decay nuclear matrix elements for both methods. We then combine the correlations with measured $2\nu\beta\beta$-decay half-lives to predict $0\nu\beta\beta$-decay matrix elements with theoretical uncertainties based on our systematic calculations. Our results include two-body currents and the short-range $0\nu\beta\beta$-decay operator.
\end{abstract}

\maketitle

\section{Introduction}
\label{sec:introduction}
Atomic nuclei can double-beta ($\beta\beta$) decay by turning two neutrons into two protons while emitting two electrons. There are two possibilities: either to emit two antineutrinos as well, as observed in a dozen nuclei~\cite{Barabash2020}, by the two-neutrino double-beta ($2\nu\beta\beta$) decay or to decay without emitting neutrinos by the yet-unobserved neutrinoless ($0\nu\beta\beta$) mode.
The latter violates lepton number conservation and the balance of matter and antimatter---only two electrons are emitted---and occurs only if neutrinos are their own antiparticles~\cite{Avignone2008,Agostini:2022zub}. Thus, detecting $0\nu\beta\beta$ decay would establish the nature of neutrinos and shed light on the matter dominance in the universe. Answering these fundamental physics questions drives ambitious worldwide $0\nu\beta\beta$-decay searches~\cite{Abe2022,Adams2021,Agostini2020,Anton2019,Alvis2019,Azzolini2019,Augier2022,Abgrall2021,Adhikari2022,Adams2021b,Albanese2021,DARWIN:2020jme}.

The $2\nu\beta\beta$- and $0\nu\beta\beta$-decay half-lives depend on well-known phase-space factors~\cite{Kotila2012} and nuclear matrix elements (NMEs)~\cite{Engel2017}. Additionally, $0\nu\beta\beta$ decay depends on a parameter encoding physics beyond the standard model of particle physics (BSM) leading to lepton-number violation. Hence, $0\nu\beta\beta$-decay NMEs are key to anticipate the reach of planned experiments in the BSM parameter space~\cite{Agostini:2021kba} and also to analyze eventual $0\nu\beta\beta$-decay signals.
For $2\nu\beta\beta$ decay, NMEs can be extracted from measured half-lives~\cite{Barabash2020}, but NMEs for $0\nu\beta\beta$ decay are poorly known: differences between state-of-the-art calculations exceed a factor three and theoretical uncertainties are mostly ignored~\cite{Engel2017,Agostini:2022zub}.

Recent theoretical works shed light on $0\nu\beta\beta$-decay NMEs. The lightest $\beta\beta$ emitter, $^{48}$Ca, has been studied with different \emph{ab initio} many-body methods~\cite{Yao:2019rck,Novario:2020dmr,Belley:2020ejd}.
First \emph{ab initio} results for heavier emitters are also available~\cite{Belley:2020ejd}, complementing more phenomenological approaches which typically describe the nuclear structure of the initial and final nuclei very well~\cite{Menendez2009,Horoi2016,Iwata2016,Menendez2018,Coraggio2020,Coraggio2022,Rodriguez2010,Barea2015}. These approaches can cover a wide range of $\beta\beta$ nuclei, but they lack the consistency which allows \emph{ab initio} methods to reproduce $\beta$-decay rates~\cite{Gysbers2019} without additional adjustments---usually known as ``quenching''. Two key aspects are to include two-body currents, which may suppress $0\nu\beta\beta$-decay NMEs as well~\cite{Menendez2011}, and additional nuclear correlations~\cite{Gysbers2019}.
On the other hand, Refs.~\cite{Cirigliano2018,Cirigliano2019} introduce a new short-range NME for $0\nu\beta\beta$ decay, with an associated hadronic coupling estimated within quantum chromodynamics (QCD)~\cite{Cirigliano:2020dmx,Cirigliano:2021qko,Richardson:2021xiu}. This leads to a significant enhancement of the $^{48}$Ca NME~\cite{Wirth:2021pij}, while the impact in heavier nuclei suggested by using approximated couplings may be similar~\cite{Jokiniemi2021}.

However, a better understanding of $0\nu\beta\beta$-decay NMEs likely requires information beyond nuclear theory to guide calculations. The structure of the initial and final $\beta\beta$-decay nuclei has received attention~\cite{Freeman12,Roberts13,Toh13,Entwisle16,Szwec16,Freeman17,Sharp19,Henderson19,Ayangeakaa19,Brown14,Rebeiro20}, as well as Gamow-Teller strengths which probe similar physics to $\beta$ decays~\cite{Ichimura06,Fujita11,Frekers18}. These properties are valuable tests of the many-body calculations, but do not show an apparent correlation with $0\nu\beta\beta$ decay. In contrast, two observables not measured so far show good correlations with $0\nu\beta\beta$-decay NMEs: double Gamow-Teller (DGT)~\cite{Shimizu2018,Yao22}, and double magnetic dipole transitions~\cite{Romeo2022}.

In this paper, we study the correlation between the NMEs of the two $\beta\beta$-decay modes for nuclei across the nuclear chart. The $2\nu\beta\beta$ and $0\nu\beta\beta$ decays share initial and final states but differ on their momentum transfers ($p$) and intermediate states. Previous studies have found a correlation between the two $\beta\beta$-decay NMEs in $^{48}$Ca~\cite{Horoi22} and a relation between their radial transition densities in all nuclei~\cite{Simkovic2011}. $2\nu\beta\beta$ decay has also been used to estimate improved $0\nu\beta\beta$-decay matrix elements~\cite{Brown:2015gsa}.
Also, $2\nu\beta\beta$-decay data is commonly used to adjust the proton-neutron quasiparticle random-phase approximation (pnQRPA) model parameters \cite{Rodin2003,Rodin2005,Faessler2008,Simkovic2013,Simkovic2018}. Here we perform systematic pnQRPA and nuclear shell model calculations, with various proton-neutron pairing strengths in the pnQRPA, and covering a wide range of nuclei and interactions in the shell model. Since $2\nu\beta\beta$-decay half-lives are known, a correlation between $\beta\beta$ NMEs can lead to $0\nu\beta\beta$-decay NMEs based on $2\nu\beta\beta$-decay data.

\section{Double-beta decay operators}
\label{sec:formalism}

The $2\nu\beta\beta$-decay half-life, to a very good approximation, depends on a single NME~\cite{Simkovic2018}:
\begin{equation}
\begin{split}
    M^{2\nu}=&\sum_{k}\frac{(0^+_{ f}||\sum_a\tau^-_a\boldsymbol{\sigma}_a||1^+_k)(1^+_k ||\sum_b\tau^-_b\boldsymbol{\sigma}_b||0^+_{i})}{(E_k-(E_i+E_f)/2)/m_e}\,,
    \end{split}
    \label{eq:2nubb}
\end{equation}
where indices $a,b$ run over all nucleons, the isospin  operator $\tau^-$ turns neutrons into protons, $\boldsymbol{\sigma}$ is the spin operator, and the denominator involves the energies $E$ of the initial ($i$), final ($f$) and each $k$th intermediate $1^+$ state. The electron mass $m_e$ makes $M^{2\nu}$ dimensionless. We solve Eq.~\eqref{eq:2nubb} directly with both the pnQRPA and shell model. For the latter framework we use the Lanczos strength function method~\cite{MPinedo}, which typically gives converged matrix elements after 50 iterations.

For $0\nu\beta\beta$ decay, we focus on the best motivated light-neutrino exchange mechanism~\cite{Agostini:2022zub}. 
The decay rate is usually written in terms of a NME with three spin structures
\begin{align}
    M^{0\nu}_{\rm L}=M_{\rm GT}^{0\nu}-M_{\rm F}^{0\nu}+M_{\rm T}^{0\nu}\;,
    \label{eq:0nbb_components}
\end{align}
called Gamow-Teller ($M_{\rm GT}^{0\nu}$), Fermi ($M_{\rm F}^{0\nu}$) and tensor ($M_{\rm T}^{0\nu}$) according to the operators $\mathcal{O}^{\rm F}_{ab}=\mathbb{I}$, $\mathcal{O}^{\rm GT}_{ab}=\boldsymbol{\sigma}_a\cdot\boldsymbol{\sigma}_b$, $\mathcal{O}^{\rm T}_{ab}=3(\boldsymbol{\sigma}_a\cdot\hat{\mathbf{r}}_{ab})(\boldsymbol{\sigma}_b\cdot\hat{\mathbf{r}}_{ab})-\boldsymbol{\sigma}_a\cdot\boldsymbol{\sigma}_b$ 
entering the definition
\begin{align}
    M^{0\nu}_K=\sum_{k,ab} (0_f^+||\mathcal{O}^K_{ab}\,\tau^-_a\tau^-_b\,H_K(r_{ab})\,f^2_{\rm SRC}(r_{ab})||0^+_i)\,,
    \label{eq:0vbb-NME}
\end{align}
where $r_{ab}$ is the distance between two nucleons.
In the pnQRPA, we sum over all intermediate states, while in the shell model we directly compute NMEs between the initial and final states in the closure approximation.
In both methods, $f_{\rm SRC}$ corrects for missing short-range correlations (SRCs) using two parametrizations~\cite{Simkovic09}. The neutrino potentials are defined as
\begin{align}
    H_K(r_{ab})=\frac{2R}{\pi g_{\rm A}^2}\int_0^{\infty}\frac{h_K\,j_{\lambda}(p\,r_{ab})\,p^ 2{\rm d}p}{\mathcal{E}_K}\;,
    \label{eq:neutrinopotential}
\end{align}
with $\mathcal{E}_K=p\,(p+E_k-(E_i+E_f)/2)$, $g_{\rm A}=1.27$ and $R=1.2A^{1/3}\,\text{fm}$ with nucleon number $A$. The spherical Bessel function $j_0$ enters all terms except the tensor where $\lambda=2$. In the shell model, we use closure with two alternative denominators $\mathcal{E}_K=p(p+1.12A^{1/2}\text{MeV})$~\cite{Haxton84} and $\mathcal{E}_K=p^2$~\cite{Cirigliano2018}. For the dominant GT term we have
\begin{align}
    h_{\rm GT}=&g^2_{\rm A}(p^2)-\frac{g_{\rm A}(p^2)g_{\rm P}(p^2)p^2}{3m_{\rm N}}+\frac{g^2_{\rm P}(p^2)p^4}{12m_{\rm N}^2} \nonumber \\
    +&\frac{g^2_{\rm M}(p^2)p^2}{6m_{\rm N}^2}\;,
    \label{eq:h_gt2}
\end{align}
and other terms are defined likewise~\cite{Engel2017}. The leading parts are proportional to the axial coupling $g_{\rm A}(p^2)$---with dipole form factor~\cite{Bernard2001}---and the pseudoscalar one $g_{\rm P}(p^2)=2m_{\rm N}g_{\rm A}(p^2)(p^2+m_{\pi}^2)^{-1}$. Here $g_{\rm M}$ is the magnetic coupling and $m_{\rm N}$, $m_{\pi}$ the nucleon and pion masses.

\begin{figure}[t]
    \centering
    \includegraphics[width=\linewidth]{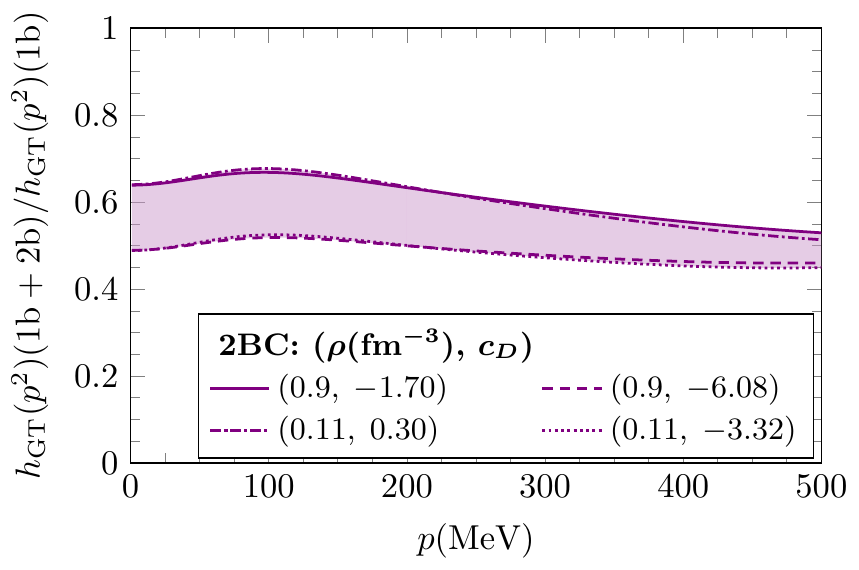}
    \caption{Relative impact of two-body currents on the function $h_{\rm GT}(p^2)$, with respect to the one-body values.}
    \label{fig:h_gt_with_2bcs}
\end{figure}

In addition to the standard shell-model and pnQRPA NMEs, we consider two additional contributions to $0\nu\beta\beta$ decay. First, we estimate the effect of two-body currents from chiral effective field theory approximated as effective one-body operators via normal ordering with respect to a spin-isospin symmetric Fermi gas reference state as in Ref.~\cite{Hoferichter2020}. The resulting current reads 
\begin{equation}
    \mathbf{J}^{\rm eff}_{i,\rm 2b}(\rho,\mathbf{p})=g_A\tau^-_i\bigg[\delta_ a(p^2)\bm{\sigma}_i+\frac{\delta _a^P(p^2)}{p^2}(\mathbf{p}\cdot\bm{\sigma}_i)\mathbf{p}\bigg],
\end{equation}
with two-body $\delta_a(p^2)$, $\delta_a^P(p^2)$ functions dependent on the Fermi-gas density $\rho$ and low-energy couplings $c_1$, $c_3$, $c_4$, $c_6$, $c_D$, for which we take the same values as in Ref.~\cite{Hoferichter2020}. This leads to the replacement 
\begin{align}
  g_{\rm A}(p^2,{\rm 2b})&\rightarrow g_{\rm A}(p^2)+\delta_a(p^2)\,,  \\
  g_{\rm P}(p^2,{\rm 2b})&\rightarrow g_{\rm P}(p^2)-\frac{2m_{\rm N}}{p^2}\,\delta_a^P(p^2)\,,
\end{align}
where the $\delta_a$, $\delta_a^P$ two-body corrections reduce $\beta$-decay NMEs by $20\%-30\%$, thus contributing to their ``quenching''. Normal-ordered currents approximate well the full two-body $\beta$-decay results~\cite{Gysbers2019}.
Figure~\ref{fig:h_gt_with_2bcs} shows the combined effect of
two-body currents in the integrand of the GT neutrino potential, $h_{\text{GT}}$. At the relevant momentum transfers $p\approx 100-200$ MeV, the suppression of the neutrino potential is rather constant, $\approx 30\%-50\%$.

\begin{figure*}[t]
    \centering
    \includegraphics[width=\linewidth]{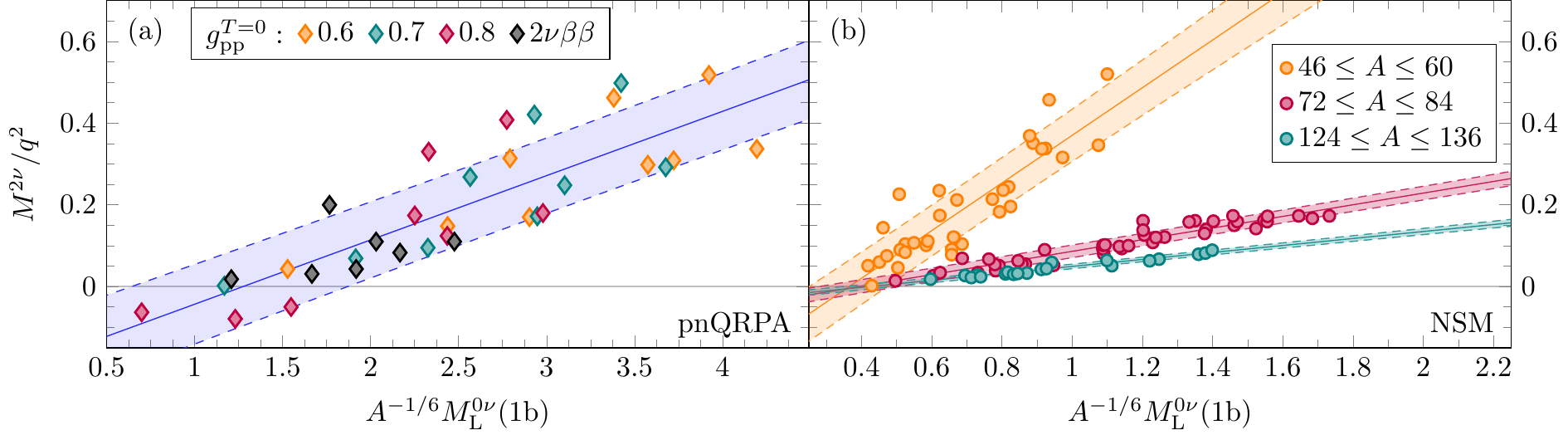}
    \caption{$2\nu\beta\beta$- ($M^{2\nu}$) vs standard $0\nu\beta\beta$-decay ($M^{0\nu}_L$) NMEs obtained with (a) pnQRPA with different isoscalar pairing $g_{\rm pp}^{T=0}$ values (adjusted to $2\nu\beta\beta$-decay data for the black diamonds) and (b) nuclear shell model (NSM) with different interactions for three regions of the nucleon number $A$. $M^{0\nu}_L$ results are multiplied by $A^{-1/6}$, and the denominator $q^2$ notes the need to quench $M^{2\nu}$ values.
    Solid and dashed lines correspond to linear fits and their 68\% CL prediction bands, respectively. }
    \label{fig:0vbb-2vbb}
\end{figure*}

Second, we also calculate the recently acknowledged short-range $0\nu\beta\beta$-decay NME~\cite{Cirigliano2018}, $M^{0\nu}_S$. This two-body term, obtained from Eq.~\eqref{eq:0vbb-NME} with $\mathcal{O}_{ab}^{\rm S}=\mathbb{I}$ but without summing over intermediate states, directly adds to the long-range part in Eq.~\eqref{eq:0nbb_components}. Because of its short-range character, it follows from $H_{\rm S}(r_{ab})$ in Eq.~\eqref{eq:neutrinopotential} with $\mathcal{E}_{\rm S}^{0\nu}=1$ and $j_0$. We use
\begin{align}
h_{\rm S}=2g_{\nu}^{\rm NN}e^{-p^2/(2\Lambda ^2)}\,,
\end{align}
with couplings $g_{\nu}^{\rm NN}$ and regulators $\Lambda$ taken from the charge-independence-breaking terms of several nuclear Hamiltonians as in Ref.~\cite{Jokiniemi2021}. This approximates the two couplings entering $\beta\beta$ decay to be equal, which for $^{48}$Ca gives a relative short-range NME contribution consistent with the {\it ab initio} result based on $g_{\nu}^{\rm NN}$ from QCD~\cite{Wirth:2021pij}.

\section{Many-body methods}
We perform nuclear shell-model calculations for the decays of a large set of nuclei in the mass range $46\leq A\leq 136$, covering three different configuration spaces with the following harmonic-oscillator single-particle orbitals---for both protons and neutrons---and isospin-symmetric interactions: i) $0f_{7/2}$, $1p_{3/2}$, $0f_{5/2}$, and $1p_{1/2}$ with the KB3G~\cite{PovesKB3G} and GXPF1B~\cite{HonmaGXPF} interactions for the decay of $^{46-58}\mbox{Ca}$, $^{50-58}\mbox{Ti}$ and $^{54-60}\mbox{Cr}$; ii) $1p_{3/2}$, $0f_{5/2}$, $\rm1p_{1/2}$ and $\rm0g_{9/2}$ with the GCN2850~\cite{Caurier08}, JUN45~\cite{HonmaJUN45} and JJ4BB~\cite{BrownJJ4BB} interactions for $^{72-76}\mbox{Ni}$, $^{74-80}\mbox{Zn}$, $^{76-82}\mbox{Ge}$ and $^{82,84}\mbox{Se}$; and iii) $1d_{5/2}$, $0g_{7/2}$, $2s_{1/2}$, $1d_{3/2}$ and $0h_{11/2}$ with the GCN5082~\cite{Caurier08} and QX~\cite{QiQX} interactions for $^{124-132}\mbox{Sn}$, $^{130-134}\mbox{Te}$ and $^{134,136}\mbox{Xe}$.
We use the shell-model codes ANTOINE~\cite{FNowacki,MPinedo} and NATHAN~\cite{MPinedo}.

In addition, we study the decays of $^{76}$Ge, $^{82}$Se, $^{96}$Zr, $^{100}$Mo, $^{116}$Cd, $^{124}$Sn, $^{128,130}$Te and $^{136}$Xe with the spherical pnQRPA method.
We use large no-core single-particle bases in a Coulomb-corrected Woods-Saxon potential~\cite{Bohr1969}
and obtain the BCS quasiparticle spectra for protons and neutrons separately. We use interactions based on the Bonn-A
potential \cite{Holinde1981}, with proton and neutron pairing fine-tuned to the empirical pairing gaps.
For the residual interaction,
we fix the particle-hole parameter
to the GT giant resonance, and the isovector particle-particle one via partial isospin-symmetry restoration~\cite{Simkovic2013}. As usual, we adjust the isoscalar particle-particle parameter to $2\nu\beta\beta$-decay half-lives. Additionally, we also explore an alternative approach and consider a range $g_{\rm pp}^{T=0}=0.6-0.8$, which gives reasonable pnQRPA NMEs for $\beta$ and $\beta\beta$ decays~\cite{Faessler2008,Pirinen2015}.

\section{Correlation Between $2\nu\beta\beta$- and $0\nu\beta\beta$-decay NMEs}

Figure~\ref{fig:0vbb-2vbb} illustrates the connection between $2\nu\beta\beta$- and $0\nu\beta\beta$-decay NMEs, where the latter only include the standard contributions in Eq.~\eqref{eq:0vbb-NME} without two-body currents or the short-range operator. To remove its mass-dependence, we multiply $M^{0\nu}$ by a factor $A^{-1/6}$~\cite{Brase2021}. Figure~\ref{fig:0vbb-2vbb} shows good linear correlations for both the pnQRPA and shell model. In the latter, the correlation depends on the nuclear mass, with a steeper slope in lighter nuclei
in the $^{48}$Ca region
and a flatter one for heavier systems
such as $^{136}$Xe.
For intermediate masses like $^{76}$Ge,
the slope is in between but closer to the one for heavy nuclei. 
The different slopes are related to the typical energies of the intermediate states which contribute the most to the $2\nu\beta\beta$-decay NME, lower for $pf$-shell nuclei and higher for heavier nuclei studied in the $sdg$ configuration space. This resembles the correlation of $0\nu\beta\beta$- and $\gamma\gamma$-decay NMEs
~\cite{Romeo2022}, even though for the latter the correlation becomes common to all nuclei with $A\geq 72$ due to the additional contribution of the orbital angular momentum operator. Nonetheless, the shell-model correlation is common for a given configuration space. The pnQRPA correlation is the same for all $\beta\beta$ emitters, and in this method the energies of the intermediates contributing most to $2\nu\beta\beta$-decay NMEs are generally higher than in the shell model~\cite{Gando19}. In both models, $2\nu\beta\beta$-decay NMEs are computed without quenching, overestimating the results. Thus we denote them in Fig.~\ref{fig:0vbb-2vbb} by $M^{2\nu}/q^2$, where $q$ is a quenching factor.
The symbols represent central values from the individual $M^{0\nu}$ results obtained with the two different SRC parametrizations and the two denominators $\mathcal{E}_K$. Table \ref{tab:NMEs} in Appendix~\ref{sec:nmes} gives a sample of our NME ranges.

We fit a linear function $M^{2\nu}/q^2=a+bA^{-1/6}M^{0\nu}$ to the central values in our NME calculations. Figure~\ref{fig:0vbb-2vbb} shows the best fits and the 68\% confidence level (CL) prediction bands. The correlation fit coefficients are $r=0.84$ for the pnQRPA and for the shell model $r=0.86$, $r=0.95$ and $r=0.97$ for the lighter, intermediate, and heavier nuclei, respectively.  
The pnQRPA NMEs obtained with smaller (larger) values of $g_{\rm pp}^{T=0}$ generally lie on the upper (lower) end of the band, while for the shell model the results calculated with different interactions are distributed rather homogeneously. Heavy nuclei are an exception, with larger (smaller) NMEs corresponding to GCN5082 (QX). The width of the prediction bands stems from the details of each NME for a given nucleus and interaction. Since our bands cover dozens of such calculations, their width can be considered as a measure of the statistical uncertainty of the results.

We correct for the overestimation of the $2\nu\beta\beta$-decay results by considering the following shell-model quenching ranges based on $\beta$- and $\beta\beta$-decay studies: $q=0.65-0.77$ for $46\leq A\leq 60$~\cite{Caurier90,Caurier94,MartinezPinedo96,Kumar:2015zuv,Kumar:2016snu}, $q=0.55-0.64$ for $72\leq A\leq 84$~\cite{Sen'kov2014,Sen'kov2014(R),Caurier12}, and $q=0.42-0.72$ for $124\leq A\leq 136$~\cite{Caurier12,Horoi16,CoelloPerez:2018ghg,Gando19}. In the pnQRPA, we assume the typical $q=0.79$ ($g_{\rm A}^{\rm eff}=1.0$) \cite{Simkovic2013,Mustonen2013,Hyvarinen2015}. It is not straightforward to quantify the quenching needed in the pnQRPA, since in the standard way of adjusting the model parameters to measured $\beta\beta$, $\beta$ or EC half-lives, $g_{\rm A}^{\rm eff}$ and $g_{\rm pp}^{T=0}$ depend on each other \cite{Faessler2008,Suhonen2013,Suhonen2014}.
The correlation band, obtained with different $g_{\rm pp}^{\rm T=0}$ values, can be considered to contain the uncertainty coming from varying the quenching.

Then, we combine the $68\%$ CL prediction bands of the linear fits with the empirical NMEs taken from $2\nu\beta\beta$-decay measurements~\cite{Barabash2020} to obtain $0\nu\beta\beta$-decay NMEs with uncertainties.
In addition, we also consider the uncertainty in the quenching needed to describe the  $2\nu\beta\beta$-decay NMEs (in the pnQRPA, we consider this to be included in the width of the band).
Finally, the total uncertainty adds quadratically the one from the width of the correlation prediction bands and the error in the NME results (see Table~\ref{tab:NMEs} in Appendix~\ref{sec:nmes}). Table~\ref{tab:NMEs_from_correlations} in Appendix~\ref{sec:nmes} gives the NME ranges for each $0\nu\beta\beta$ decay.

\begin{figure}[t]
    \centering
    \includegraphics[width=\linewidth]{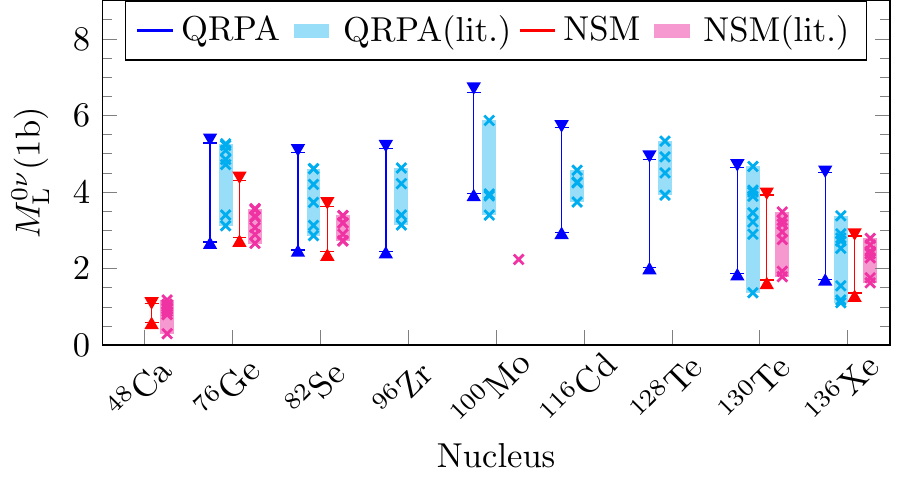}
    \caption{Standard $0\nu\beta\beta$-decay NMEs obtained from the correlations in Fig.~\ref{fig:0vbb-2vbb}. The narrow error bars come from the 68\% CL bands of the linear fits, while the wide ones also contain uncertainties in the NME calculations.  Bands (crosses) show the literature NME ranges (individual values), shell model (NSM) in red~\cite{Horoi2016,Iwata2016,Menendez2018,Coraggio2020,Coraggio2022,Jokiniemi2021}, QRPA in blue~\cite{Mustonen2013,Hyvarinen2015,Simkovic2018,Fang2018,Terasaki2020,Jokiniemi2021}.}
    \label{fig:0vbb(1b)-correlation-estimates}
\end{figure}

Figure~\ref{fig:0vbb(1b)-correlation-estimates} compares the shell-model (red) and pnQRPA (blue) $0\nu\beta\beta$-decay NMEs obtained from the correlation and $2\nu\beta\beta$-decay data.
The narrow error bars (marked by horizontal lines) are derived from the 68\% CL prediction bands of the fits---and the much smaller uncertainties in the empirical $2\nu\beta\beta$-decay NMEs---while the wider bars (marked by triangles) add these errors quadratically to the ones of the individual NME results. The latter contribution is typically the only one considered when giving theoretical $0\nu\beta\beta$-decay NME uncertainties, and it is always much smaller than the error associated with the NME correlation.
While pnQRPA NMEs are larger than the shell-model ones, Fig.~\ref{fig:0vbb(1b)-correlation-estimates} shows that considering error bars both methods are consistent. The uncertainty is relatively larger in the pnQRPA because of the smaller correlation coefficient than in the shell model.
The shell-model error bars are larger for $^{130}$Te and $^{136}$Xe due to the more uncertain quenching for heavy nuclei.

Figure~\ref{fig:0vbb(1b)-correlation-estimates} also compares our NMEs derived using the $2\nu\beta\beta-0\nu\beta\beta$ correlation with previous shell-model~\cite{Horoi2016,Iwata2016,Menendez2018,Coraggio2020,Coraggio2022,Jokiniemi2021} and pnQRPA~\cite{Mustonen2013,Hyvarinen2015,Simkovic2018,Fang2018,Terasaki2020,Jokiniemi2021} results. There is generally a good agreement: our error bars cover the range of earlier results with only a few exceptions. For  $^{76}$Ge, the shell-model NME is somewhat larger than in previous works which underestimate this nucleus' $2\nu\beta\beta$-decay NME. In the pnQRPA, the $^{100}$Mo NME is also larger mainly because its exceptionally short half-life affects all intermediate states through the correlation, but mostly $1^+$ states---sensitive to $g^{T=0}_{\text{pp}}$---in previous works.
We note that while the literature band comprises at most a handful different calculations, our uncertainty obtained from the correlation includes information from systematic results for tens of nuclei using several interactions. Moreover, our shell-model $M_L^{0\nu}$ for $^{48}$Ca is in excellent agreement with the statistical analysis of Ref.~\cite{Horoi22} based on 20\,000 calculations each performed with a different variation of three independent shell-model interactions.

\begin{figure}[t]
    \centering
    \includegraphics[width=\linewidth]{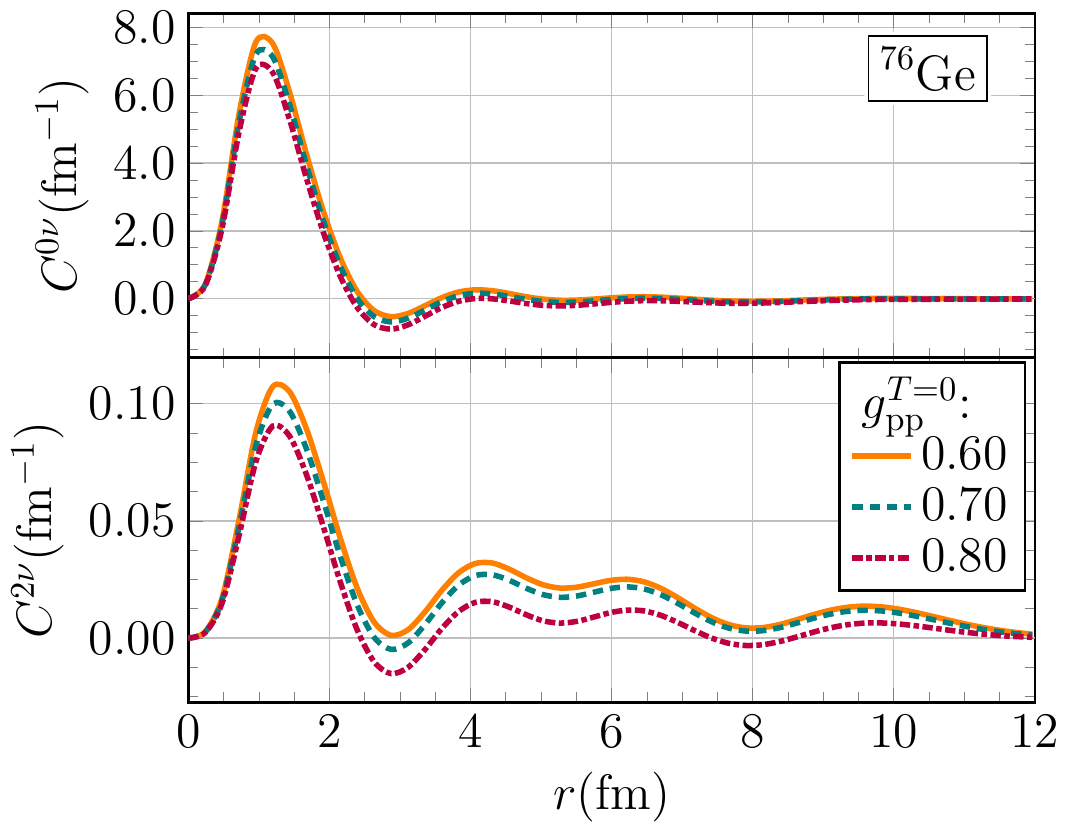}
    \caption{Radial distributions of the $0\nu\beta\beta$- (top panel) and $2\nu\beta\beta$-decay (bottom) NMEs of $^{76}$Ge obtained with the pnQRPA. The $2\nu\beta\beta$-decay densities are obtained using the correct energy denominator for the $1^+$ states and the closure energy $E_{\rm GTGR}=11.4$ MeV for the rest.}
    \label{fig:76Ge-0vbb(r)-2vbb(r)_edenGT}
\end{figure}

\section{Radial distributions and multipole decomposition}
\label{sec:radial}
In order to understand the origin of the correlation between $M^{2\nu}$ and $M^{0\nu}$,
Fig.~\ref{fig:76Ge-0vbb(r)-2vbb(r)_edenGT} shows the NME radial distributions for $^{76}$Ge obtained with the pnQRPA. The radial distributions for other isotopes are qualitatively similar. They satisfy
\begin{align}
    M^{0\nu}_{\rm L}({\rm 1b})=&\int_0^{\infty}C^{0\nu}(r){\rm d}r\;,\\
    M^{2\nu}=&\int_0^{\infty}C^{2\nu}(r){\rm d}r\;,
    \label{eq:M2nu_int_C2nu}
\end{align}
and are defined as
\begin{align}
\label{eq:0nu_rad_dens}
&C^{0\nu}(r)=C^{0\nu}_{\rm GT}(r)-C^{0\nu}_{\rm F}(r)+C^{0\nu}_{\rm T}(r)\;, \\
    &C^{0\nu}_{\rm K}(r)= \nonumber \\ &\sum_{k,ab}(0^+_f||\mathcal{O}^K_{ab}\tau^-_a\tau^-_bH_K(r_{ab})f_{\rm SRC}^2(r_{ab})\delta(r-r_{ab})||0^+_i)\;,
\end{align}
and 
\begin{align}
 \label{eq:2nu_rad_dens}
    C^{2\nu}(r)&=\sum_{k,ab}\frac{(0^+_f||\boldsymbol{\sigma}_a\cdot\boldsymbol{\sigma}_b\tau^-_a\tau^-_b\delta(r-r_{ab})||0^+_i)}{D(J^{\pi}_k)/m_e}\;, \\
   D(J^{\pi}_k)&=\begin{cases}
   E_k-(E_i+E_f)/2\;,\quad \text{if }J^{\pi}=1^+\\
   E_{\rm GTGR}=11.4\;\text{MeV, }\quad \text{if }J^{\pi}\neq 1^+\;.  
   \end{cases}
\end{align}
While $M^{2\nu}$ gets contributions only from $1^+$ intermediate states, all possible $J^{\pi}$ multipoles play a role in $C^{2\nu}$ due to the radial dependence of $\delta(r-r_{ab})$~\cite{Simkovic2011}. Nonetheless, when integrated as in Eq.~\eqref{eq:M2nu_int_C2nu}, $J^{\pi}\neq 1^+$ contributions vanish. Since $M^{2\nu}$ does not depend on these multipoles, it is not clear how to deal with their energy denominator in Eq.~\eqref{eq:2nu_rad_dens}. Here we take as closure energy $E_{\rm GTGR}=11.4$ MeV, to which the particle-hole parameter of the pnQRPA is adjusted \cite{Jokiniemi2018}, and the correct energy denominator for $1^+$ intermediate states. This guarantees that the integral of the distribution leads to the correct value for $M^{2\nu}$. However, following Eq.~\eqref{eq:2nu_rad_dens}, the shape of $C^{2\nu}$ is sensitive to the closure energy used, so that it comes with some uncertainty. For example, varying the $D(J^{\pi}\neq 1^+)$ by $\pm1$ MeV decreases (increases) the short-range contribution below $r\lesssim 3$ fm by some 5\% and increases (decreases) the long-range contribution accordingly. However, qualitatively the distribution remains similar.

Figure~\ref{fig:76Ge-0vbb(r)-2vbb(r)_edenGT} shows that long-range contributions can play an important role in $2\nu\beta\beta$ decay, especially with small proton-neutron pairing values: for $g_{\rm pp}^{T=0}=0.6$ the contribution from $r\gtrsim3$~fm is as much as 50\%, with the same sign as the short-range peak. However, for larger $g_{\rm pp}^{T=0}$ values the long-range part is less important, and partially cancels with the intermediate-range part from $2.5~{\rm fm} \lesssim r\lesssim3.5~{\rm fm}$ part, so that the NME is effectively driven by the short distances. In contrast, the top panel shows that $0\nu\beta\beta$ decay is always dominated by shorter distances, being less sensitive to the value of the $g_{\rm pp}^{T=0}$ parameter.

These radial distributions $C^{2\nu}$ resemble the ones of DGT transitions, which are well correlated with $0\nu\beta\beta$-decay NMEs in the shell model, EDF theory, and the IBM~\cite{Shimizu2018}, as well as in {\it ab initio}~\cite{Yao22} calculations. For small $g_{\rm pp}^{T=0}$ values, the significant long-range contribution enhancing the NME is similar to {\it ab initio} isospin-conserving DGT transitions, which show a strong correlation with $0\nu\beta\beta$-decay NMEs. On the other hand, for smaller $g_{\rm pp}^{T=0}$ values the distributions are comparable to shell-model DGT ones, where the short-range behaviour leads to a good correlation with $0\nu\beta\beta$ decay~\cite{Anderson2010,Bogner2012}. We also note that our distributions $C^{2\nu}$ do not show a dominant long-range part cancelling the short-range peak, a shape that has been shown to degrade the correlation~\cite{Yao22}. Hence, the pnQRPA radial distributions show features similar to previous studies which also observe good correlations with $0\nu\beta\beta$-decay NMEs.  We are not able to study these distributions with the shell model because we do not compute contributions from each $J^{\pi}$ multipole. 

\setlength{\tabcolsep}{8.5pt}
\begin{table}[t!]
\caption{Correlation coefficients $r$ for the linear relations between $M^{2\nu}$ and the lowest multipole components $M^{0\nu}(J^{\pi})$. The average share of $M^{0\nu}$ for each multipole is also shown.}
    \label{tab:correlation_coeffs_multipoles}
    \centering
    \begin{ruledtabular}
    \begin{tabular}{ccc}
    $J^{\pi}$ &$r$ &Average $M^{0\nu}(J^{\pi})/M^{0\nu}$(\%)\\
    \midrule
         $0^+$& 0.26 & 0.6\\
         $1^+$& 0.79 &12.5 \\
         $2^+$& 0.73 &9.3 \\
         $3^+$& 0.81 &7.6 \\
         $0^-$& 0.77 &0.7 \\
         $1^-$& 0.85 &11.5 \\
         $2^-$& 0.73 &9.3 \\
         $3^-$& 0.65 &7.5 \\
    \end{tabular}
    \end{ruledtabular}
\end{table}

Furthermore, we study the correlation of $M^{2\nu}$ and the different $J^{\pi}$ multipoles contributing to $M^{0\nu}$:
\begin{equation}
    M^{0\nu}_{\rm L}({\rm 1b})=\sum_{J^{\pi}}M^{0\nu}(J^{\pi})\;.
\end{equation}
We explore these additional correlations with the pnQRPA because our shell-model calculations use the closure approximation. Table \ref{tab:correlation_coeffs_multipoles} lists the correlation coefficients together with the average share to the total $M^{0\nu}$ of the lowest multipoles. For the $1^+,\;2^+,\;3^+,\;0^-,\;1^-$ and $2^-$ multipoles, which altogether constitute on average 51\% of the total NME, we find good correlations with $r>0.70$, and for $3^-$, which gives $\approx 7.5\%$ of the NME, the coefficient is still $r=0.65$. Only the $0^+$ part does not seem to be correlated with $M^{2\nu}$, yet its contribution to $M^{0\nu}$ is negligible. Hence, in the pnQRPA $M^{2\nu}$ is not only correlated with $M^{0\nu}$, but also with its most important multipole components. This could also explain why in other many-body methods DGT NMEs are correlated with $M^{0\nu}$, even if the former only receive contributions from $1^+$ intermediate states just like $M^{2\nu}$.

\section{Two-body currents and short-range $0\nu\beta\beta$-decay NME}

The effect of two-body currents on $0\nu\beta\beta$-decay NMEs is similar in the shell model and pnQRPA: NMEs decrease by $25\%-45\%$. The range is mainly driven by the uncertainties in $\delta_a$, $\delta_a^P$. This reduction is somewhat larger than in earlier studies~\cite{Menendez2011,Engel2014} which neglect pion-pole diagrams~\cite{Klos:2013rwa}. In contrast to Ref.~\cite{Menendez2011}, the effect of two-body currents with these additional contributions is fairly constant at $p\approx100-250$~MeV, relevant for $0\nu\beta\beta$ decay, see Fig.~\ref{fig:h_gt_with_2bcs}. Since two-body currents impact all nuclei rather uniformly, we also find a good linear correlation between $2\nu\beta\beta$- and $0\nu\beta\beta$-decay NMEs in this case.
Table \ref{tab:correlations} presents the parameters of all NME correlations, where $46\leq A\leq60$ nuclei are denoted by $pf$, $72\leq A\leq 84$ isotopes by $pfg$ and $124\leq A\leq 136$ nuclei by $sdg$.
In particular, Table \ref{tab:correlations} shows that the correlation coefficients remain practically unchanged when two-body currents are included.

\begin{figure}[t]
    \centering
    \includegraphics[width=\linewidth]{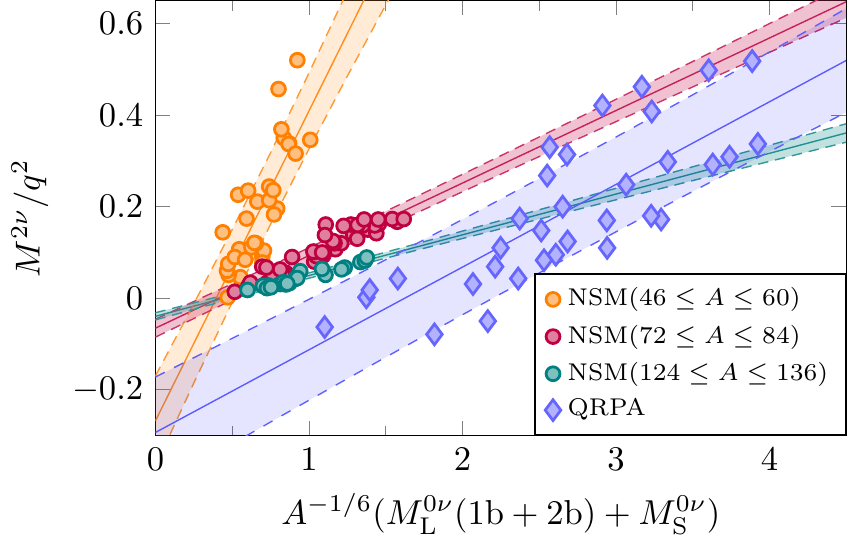}
    \caption{$0\nu\beta\beta$- vs $2\nu\beta\beta$-decay NMEs and linear fits with 68\% CL prediction bands for the shell model (NSM, circles) and pnQRPA (diamonds, for all $g_{\rm pp}^{T=0}$ values in Fig.~\ref{fig:0vbb-2vbb}). For $0\nu\beta\beta$ decay, results include two-body currents and short-range NMEs.}
    \label{fig:0vbb-2vbb-2BCS-MS-NSM-vs-QRPA}
\end{figure}

Finally, we add the short-range operator into $0\nu\beta\beta$-decay NMEs. In the pnQRPA, this term typically amounts to some $30\%-80\%$ of the one-body $M^{0\nu}_{\rm L}$ value, and in the shell model this fraction is about $15\%-50\%$. Individual uncertainties are now larger, dominated by the short-range coupling $g_{\nu}^{\rm NN}$. Figure~\ref{fig:0vbb-2vbb-2BCS-MS-NSM-vs-QRPA} shows the corresponding correlations between the $2\nu\beta\beta$- and $0\nu\beta\beta$-decay NMEs, with symbols denoting central NME values.
The pnQRPA results include all $g_{\rm pp}^{T=0}$ values shown in Fig.~\ref{fig:0vbb-2vbb}.
Here the correlation coefficients become $r=0.80$ in the pnQRPA and $0.81\leq r \leq 0.97$ in the shell model (see Table~\ref{tab:correlations}), smaller than in previous cases because the short-range term has Fermi spin structure, which does not contribute to $2\nu\beta\beta$ decay. Figure~\ref{fig:0vbb-2vbb-2BCS-MS-NSM-vs-QRPA} also highlights that the slope of the pnQRPA correlation is similar to the shell-model one for $^{76}$Ge, and not very different to the one for $^{136}$Xe---note that Fig.~\ref{fig:0vbb-2vbb-2BCS-MS-NSM-vs-QRPA} does not show pnQRPA results for nuclei as light as $^{48}$Ca. However, since the pnQRPA generally predicts larger $M^{0\nu}$ values than the shell model, its correlation is shifted to the right.

\begin{table}[t]
    \caption{Parameters and correlation coefficient $r$ of the linear fits $M^{2\nu}/q^2=a+bA^{-1/6}M^{0\nu}$. The first column indicates whether the results include only $M^{0\nu}_{\rm L}$ (L) or both $M^{0\nu}_{\rm L}$ and $M^{0\nu}_{\rm S}$ (L+S) and whether they cover only one-body currents (1b) or both one- and two-body currents (2b). The correlations are shown in Figs.~\ref{fig:0vbb-2vbb}, ~\ref{fig:0vbb-2vbb-2BCS-MS-NSM-vs-QRPA}, ~\ref{fig:intermediate_steps} 
    and \ref{fig:intermediate_steps_SM} for the pnQRPA and shell model (NSM).}
    \centering
   \begin{ruledtabular}
    \begin{tabular}{ccccc}
    $M^{0\nu}$ &Model &$a$ &$b$ & $r$ \\
    \midrule
         L,1b &pnQRPA &-0.201 &0.157 &0.84  \\ 
         L,2b &pnQRPA &-0.227 &0.256 &0.84 \\ 
         L+S,1b &pnQRPA &-0.263 &0.128 &0.82\\ 
         L+S,2b &pnQRPA &-0.293 &0.181 &0.80 \\
         \hline
         L,1b &NSM [pf] &-0.215 &0.589 &0.86 \\ 
         L,1b &NSM [pfg] &-0.056 &0.143 &0.95 \\ 
         L,1b &NSM [sdg] &-0.036 &0.085  &0.97 \\ 
         \hline
         L,2b &NSM [pf] &-0.256 &0.479 &0.84 \\
         L,2b &NSM [pfg] &-0.063 &0.112 &0.94 \\
         L,2b &NSM [sdg] &-0.039 &0.065 &0.97 \\
         \hline
         L+S,1b &NSM [pf] &-0.215 &0.939 &0.84\\
         L+S,1b &NSM [pfg] &-0.056 &0.226 &0.95 \\
         L+S,1b &NSM [gds] &-0.036 &0.134 &0.97 \\
         \hline
         L+S,2b &NSM [pf] &-0.268 &0.677 &0.81 \\
         L+S,2b &NSM [pfg] &-0.065 &0.159 &0.94 \\
         L+S,2b &NSM [sdg] &-0.040 &0.089 &0.97 \\
    \end{tabular}
    \end{ruledtabular}
    \label{tab:correlations}
\end{table}

\begin{figure*}
    \centering
    \includegraphics[scale=0.84]{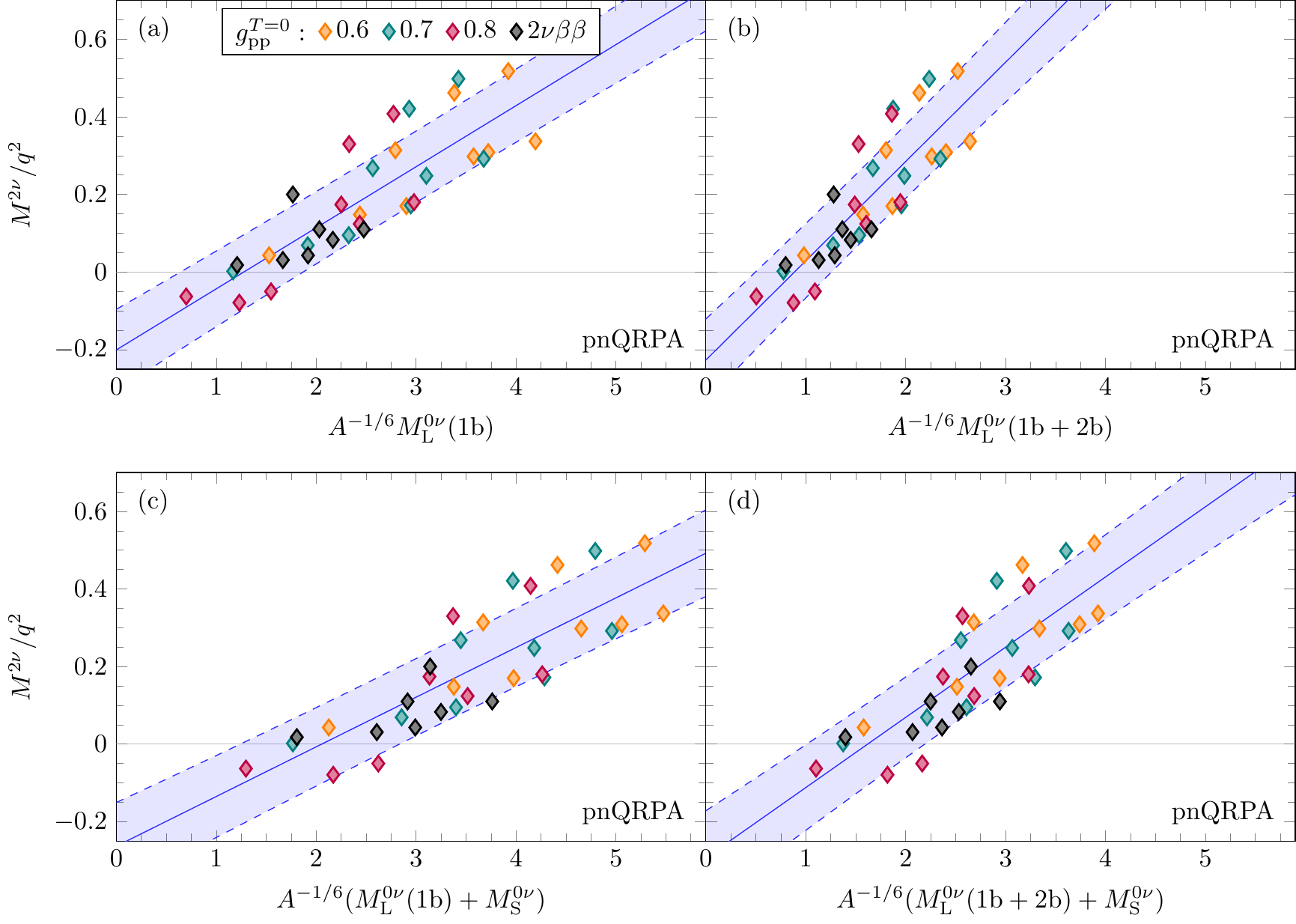}
    \caption{Correlation between pnQRPA $M^{2\nu}$ and $M^{0\nu}$ NMEs with (panels b, d) or without (panel a, c) two-body currents and with (panels c, d) or without (panels a, b) the short-range $0\nu\beta\beta$-decay term.}
    \label{fig:intermediate_steps}
\end{figure*}

\begin{figure*}
    \centering
    \includegraphics[scale=0.84]{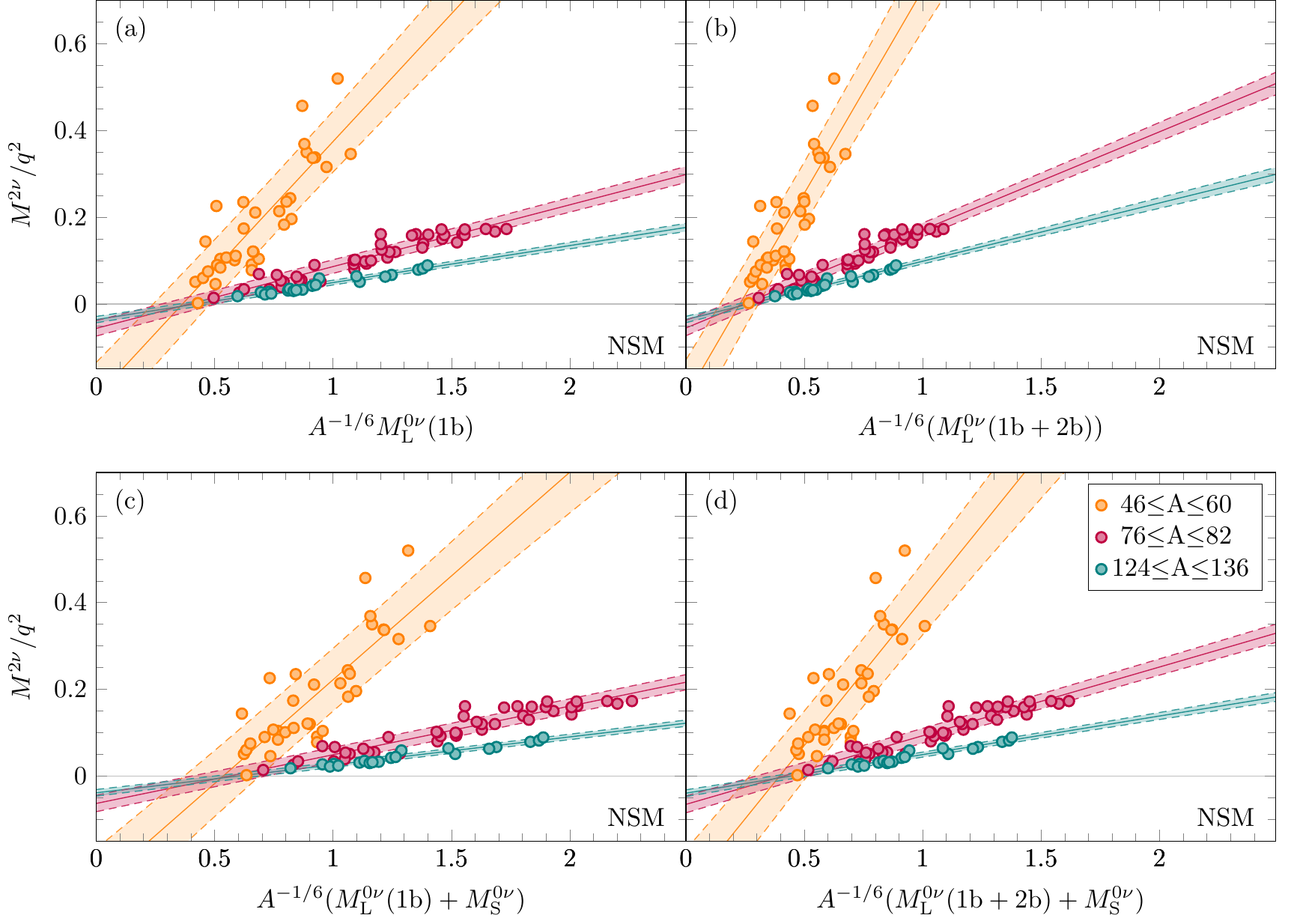}
    \caption{Same as Fig.~\ref{fig:intermediate_steps} but for the nuclear shell model (NSM).}
    \label{fig:intermediate_steps_SM}
\end{figure*}

Figures~\ref{fig:intermediate_steps} (for the pnQRPA) and \ref{fig:intermediate_steps_SM} (for the nuclear shell model) show the different correlations we obtain between the $2\nu\beta\beta$- and $0\nu\beta\beta$-decay NMEs in terms of which components of the $0\nu\beta\beta$-decay NMEs we consider. For the sake of a better comparison, panels (a) in Figs.~\ref{fig:intermediate_steps} and \ref{fig:intermediate_steps_SM} show the same correlations in Fig.~\ref{fig:0vbb-2vbb}.
Since adding the effective two-body currents results in relative reduction of $0\nu\beta\beta$-decay NMEs by some $25\%-45\%$ for both many-body methods, the correlations in panels (b) in Figs.~\ref{fig:intermediate_steps} and \ref{fig:intermediate_steps_SM} are shifted towards negative $x$-axis and the slopes increase. On the other hand, since the short-range NMEs enhance the $0\nu\beta\beta$-decay NMEs, adding this contribution to the $0\nu\beta\beta$-decay NMEs shifts the correlations in panels (c) towards positive $x$-axis and decreases the slopes. Hence the two effects tend to balance each other, and once both two-body currents and the short-range $0\nu\beta\beta$-decay NME are added in panels (d), the correlations resemble those obtained with $M^{0\nu}_ {\rm L}({\rm 1b})$ only. Table \ref{tab:correlations} clearly highlights that the effects of two-body currents and the short-range NME partially cancel, and the best linear fits of the correlations of $M^{2\nu}$ with $M^{0\nu}_ {\rm L}({\rm 1b})$ and $M^{0\nu}_ {\rm L}({\rm 1b+2b})+M^{0\nu}_ {\rm S}$  are relatively similar.

\begin{figure}[t]
    \centering
    \includegraphics[width=\linewidth]{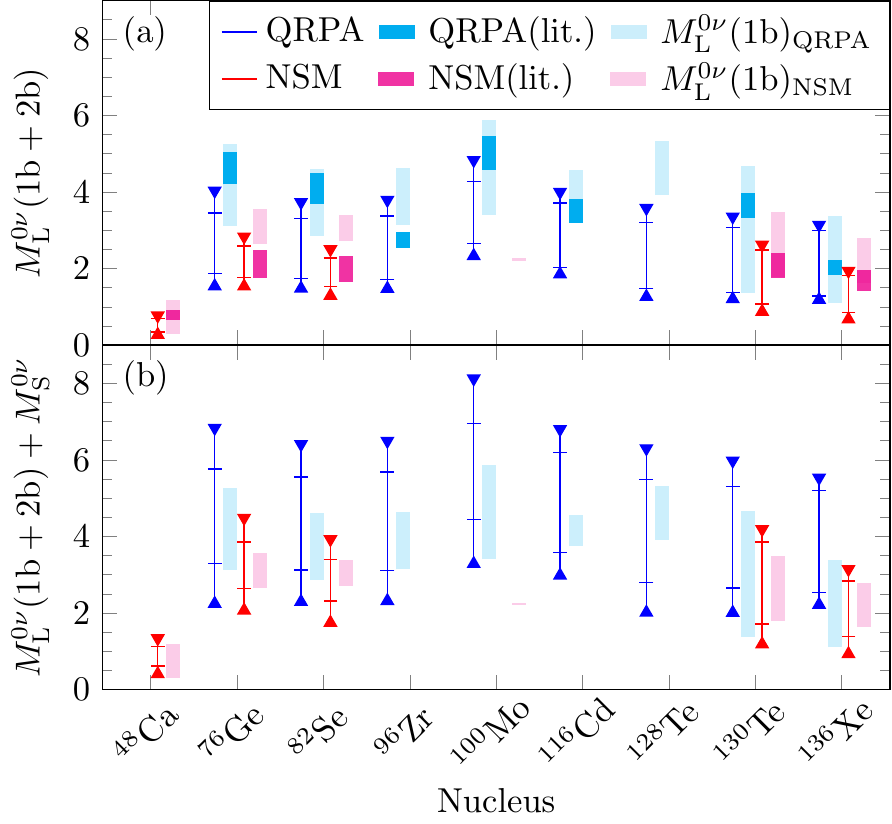}
    \caption{$0\nu\beta\beta$-decay NMEs with error bars derived from correlation fits as in Fig. \ref{fig:0vbb(1b)-correlation-estimates}. (a) NMEs with two-body currents compared to Refs.~\cite{Engel2014,Menendez2011} (dark bands). (b) NMEs with two-body currents and the short-range term. For comparison, light bands show the literature bands of Fig.~\ref{fig:0vbb(1b)-correlation-estimates}.}
    \label{fig:0vbb(1b+2b)+MS-correlation-estimates}
\end{figure}

Figure~\ref{fig:0vbb(1b+2b)+MS-correlation-estimates} shows $0\nu\beta\beta$-decay NMEs with two-body currents and short-range NMEs derived from the correlations and  $2\nu\beta\beta$-decay data. Panel (a) shows that two-body currents reduce the NMEs (light bands correspond to the bands in Fig.~\ref{fig:0vbb(1b)-correlation-estimates} for reference). In fact, especially pnQRPA but also shell-model NMEs with two-body currents are notably smaller than in previous works~\cite{Engel2014,Menendez2011} (shown as dark bands) mostly due to the more complete currents considered here. The total error bars are wider than in Fig.~\ref{fig:0vbb(1b)-correlation-estimates} because of the uncertainties in $\delta_a$, $\delta_a^P$. Our shell-model $M_L^{0\nu}({\rm 1b+2b})$ NMEs are in good agreement with {\it ab initio} results for $^{48}$Ca~\cite{Yao:2019rck,Belley:2020ejd,Novario:2020dmr} and $^{76}$Ge~\cite{Belley:2020ejd} within uncertainties, and for $^{82}$Se our error bar is just above the {\it ab initio} value~\cite{Belley:2020ejd}. This suggests that $\delta_a$, $\delta_a^P$ effectively capture part of the missing many-body correlations---note that {\it ab initio} $0\nu\beta\beta$-decay NMEs do not include two-body currents yet. Further, our shell-model $M_L^{0\nu}({\rm 1b+2b})$ NMEs are consistent---with lower central values and larger uncertainties---with Ref.~\cite{Weiss:2021rig}, which follows a different approach for adding correlations into the shell-model framework.

Figure~\ref{fig:0vbb(1b+2b)+MS-correlation-estimates} (b) shows that when we include the short-range operator, $0\nu\beta\beta$-decay NMEs obtained from the correlation and $2\nu\beta\beta$-decay data become comparable with the standard ones in Fig.~\ref{fig:0vbb(1b)-correlation-estimates}
(again, light bands serve as a reference). However, error bars become notably larger due to the sizeable uncertainties especially in the short-range coupling $g_{\nu}^{\rm NN}$, which are comparable to the uncertainties from the NME correlation.

\section{Summary}

We perform shell-model and pnQRPA calculations for several tens of $\beta\beta$ decays and nuclear interactions and observe good linear correlations between $2\nu\beta\beta$- and $0\nu\beta\beta$-decay NMEs.
We also find good correlations when including two-body currents and the short-range operator into $0\nu\beta\beta$ decay, even though in these cases the uncertainty of the NME calculations increases driven by uncertainties in the couplings associated with these contributions.
Using the correlations and measured $2\nu\beta\beta$ decays, we obtain $0\nu\beta\beta$-decay NMEs with theoretical uncertainties based on systematic calculations following the same correlation, rather than individual NME results. Our nuclear matrix elements are generally in good agreement with previous shell-model and pnQRPA studies. While the theoretical uncertainties derived in this work can be larger than the spread of previous shell-model and pnQRPA matrix elements, we stress that our strategy is based on correlations built from dozens of decays, and may therefore be considered more reliable than individual NME calculations.
Many-body approaches able to compute $2\nu\beta\beta$-decay NMEs~\cite{Barea2015,Novario:2020dmr,Belley:2020ejd,Jokiniemi:2022yfr} could pursue similar strategies to predict $0\nu\beta\beta$-decay NMEs with theoretical uncertainties.

\section*{Acknowledgements}
This work was supported by the Finnish Cultural Foundation grant No. 00210067, Arthur B. McDonald Canadian Astroparticle Physics Research Institute, and by the ``Ram\'on y Cajal'' program with grant RYC-2017-22781, and grants CEX2019-000918-M, PID2020-118758GB-I00 and RTI2018-095979-B-C41 funded by MCIN/AEI/10.13039/501100011033 and by ``ESF Investing in your future.'' TRIUMF receives federal funding via a contribution agreement with the National Research Council of Canada.

\bibliography{0vbb-2vbb-bib}

\appendix

\section{$0\nu\beta\beta$-decay NME values}
\label{sec:nmes}

Table \ref{tab:NMEs} collects a sample of the $0\nu\beta\beta$-decay NME ranges we have obtained for $\beta\beta$ emitters using the pnQRPA and the shell model---the latter calculations involve many other $\beta\beta$ decays as well. The results cover all calculations including or not two-body currents and the short-range NME.

Our final $0\nu\beta\beta$-decay NME ranges obtained using the best linear fits to each correlation, for any combination of $0\nu\beta\beta$-decay NMEs---with and without two-body currents or the short-range $0\nu\beta\beta$-decay NME---are summarized in Table~\ref{tab:NMEs_from_correlations}.

\begin{table*}
\centering
\caption{Sample of calculated $0\nu\beta\beta$-decay NMEs for $\beta\beta$ emitters, just the standard NME $M^{0\nu}_L$(1b) and including two-body currents $M^{0\nu}_L$(1b+2b) and/or the short-range $M^{0\nu}_S$. The pnQRPA values listed here use $g_{\rm pp}^{T=0}$ adjusted to $2\nu\beta\beta$-decay data, except for $^{96}$Zr for which $g_{\rm pp}^{T=0}=0.8$. The NME ranges cover different SRCs and denominators $\mathcal{E}_K$ for the shell model (NSM).}
\begin{ruledtabular}
\begin{tabular}{llcccc}
Nucleus & Model & $M_{\rm L}^{0\nu}(\rm{1b})$ &$M_{\rm L}^{0\nu}(\rm{1b})+M_{\rm{S}}^{0\nu}$ &$M_{\rm L}^{0\nu}(\rm{1b+2b})$& $M_{\rm{L}}^{0\nu}({\rm 1b+2b})+M_{\rm{S}}^{0\nu}$\\
\hline
\multirow{2}{*}{$^{48}$Ca} & NSM (KB3G) &$0.87 - 1.05$ &$1.10-1.71$ &$0.50 - 0.73$ &$0.73 - 1.39$\\
& NSM (GXPF1B) &$0.72 - 0.87$ &$0.92-1.46$ &$0.42 - 0.61$ &$0.62 - 1.19$\\
\hline
\multirow{5}{*}{$^{76}$Ge} & \multirow{2}{*}{pnQRPA} & \multirow{2}{*}{$4.83-5.36$} & \multirow{2}{*}{$6.32-9.16$}& \multirow{2}{*}{$2.97-3.85$} & \multirow{2}{*}{$4.46-7.65$}\\
& & & & \\
& NSM (GCN2850) &$2.85 - 3.52$ &$3.36-5.00$ &$1.58 - 2.40$ &$2.10 - 3.89$\\
& NSM (JUN45) &$3.11 - 3.82$ &$3.66-5.40$ &$1.74 - 2.62$ &$2.28 - 4.20$\\
& NSM (JJ4BB) &$2.86 - 3.54$ &$3.34-4.92$ &$1.60 - 2.42$ &$2.07 - 3.80$\\
\hline
\multirow{5}{*}{$^{82}$Se} & \multirow{2}{*}{pnQRPA} & \multirow{2}{*}{$4.30-4.73$} & \multirow{2}{*}{$5.57-7.97$}& \multirow{2}{*}{$2.64-3.40$} & \multirow{2}{*}{$3.91-6.64$}\\
& & & & \\
& NSM (GCN2850) &$2.71 - 3.37$ &$3.19-4.75$ &$1.51 - 2.30$ &$1.99 - 3.68$\\
& NSM (JUN45) &$2.91 - 3.58$ &$3.41-5.04$ &$1.62 - 2.46$ &$2.13 - 3.92$\\
& NSM (JJ4BB) &$2.47 - 3.08$ &$2.89-4.29$ &$1.38 - 2.11$ &$1.80 - 3.31$\\
\hline
\multirow{2}{*}{$^{96}$Zr} & \multirow{2}{*}{pnQRPA} & \multirow{2}{*}{$4.75-5.22$} & \multirow{2}{*}{$5.99-8.43$}& \multirow{2}{*}{$2.84-3.70$} & \multirow{2}{*}{$4.08-6.91$}\\
& & & & \\
\hline
\multirow{2}{*}{$^{100}$Mo} & \multirow{2}{*}{pnQRPA} & \multirow{2}{*}{$3.52-4.09$} & \multirow{2}{*}{$5.18-8.35$}& \multirow{2}{*}{$2.41-3.10$} & \multirow{2}{*}{$4.07-7.36$}\\
& & & & \\
\hline
\multirow{2}{*}{$^{116}$Cd} & \multirow{2}{*}{pnQRPA} & \multirow{2}{*}{$4.31-4.66$} & \multirow{2}{*}{$5.41-7.46$}& \multirow{2}{*}{$2.66-3.37$} & \multirow{2}{*}{$3.76-6.17$}\\
& & & & \\
\hline
\multirow{2}{*}{$^{128}$Te} & \multirow{2}{*}{pnQRPA} & \multirow{2}{*}{$4.09-4.52$} & \multirow{2}{*}{$5.46-7.97$}& \multirow{2}{*}{$2.51-3.28$} & \multirow{2}{*}{$3.88-6.73$}\\
& & & & \\
\hline
\multirow{4}{*}{$^{130}$Te} & \multirow{2}{*}{pnQRPA} & \multirow{2}{*}{$3.52-3.98$} & \multirow{2}{*}{$4.70-7.03$} & \multirow{2}{*}{$2.20-2.88$} & \multirow{2}{*}{$3.38-5.93$}\\
& & & & \\
& NSM (GCN5082) &$2.75 - 3.46$ &$3.32-5.10$ &$1.56 - 2.38$ &$2.12 - 4.02$\\
& NSM (QX) &$1.64 - 2.04$ &$2.00-3.09$ &$0.93 - 1.40$ &$1.30 - 2.46$\\
\hline
\multirow{4}{*}{$^{136}$Xe} & \multirow{2}{*}{pnQRPA} & \multirow{2}{*}{$2.59-2.89$} & \multirow{2}{*}{$3.35-4.84$}& \multirow{2}{*}{$1.56-2.06$} & \multirow{2}{*}{$2.32-4.01$}\\
& & & & \\
& NSM (GCN5082) &$2.21 - 2.78$ &$2.66-4.08$ &$1.25 - 1.91$ &$1.70 - 3.22$\\
& NSM (QX) &$1.50 - 1.86$ &$1.82-2.81$ &$0.85 - 1.28$ &$1.18 - 2.23$\\
\end{tabular}
\end{ruledtabular}
\label{tab:NMEs}
\end{table*}

\begin{table*}
\centering
\caption{$0\nu\beta\beta$-decay NMEs like in Table~\ref{tab:NMEs} but obtained from the NME correlation fits and $2\nu\beta\beta$-decay data. In the shell model (NSM) we use quenching factors $q=0.65-0.77$ for $^{48}$Ca, $q=0.55-0.64$ for $^{76}$Ge, $^{82}$Se and $q=0.42-0.72$ for $^{130}$Te, $^{136}$Xe. In the pnQRPA, we use $q=0.79$ for all the nuclei. The NME ranges combine the errors derived from the fits and the NME calculations added quadratically.}
\begin{ruledtabular}
\begin{tabular}{llcccc}
Nucleus & Model & $M_{\rm L}^{0\nu}(\rm{1b})$ &$M_{\rm{L}}^{0\nu}({\rm 1b})+M_{\rm{S}}^{0\nu}$  &$M_{\rm L}^{0\nu}(\rm{1b+2b})$ & $M_{\rm{L}}^{0\nu}({\rm 1b+2b})+M_{\rm{S}}^{0\nu}$\\
\hline
\multirow{2}{*}{$^{48}$Ca} &  \multirow{2}{*}{NSM} & \multirow{2}{*}{$0.58-1.10$}& \multirow{2}{*}{$0.62-1.66$} & \multirow{2}{*}{$0.32-0.73$} &\multirow{2}{*}{$0.45-1.29$}\\
& & & & \\
\hline
\multirow{3}{*}{$^{76}$Ge} & \multirow{2}{*}{pnQRPA} & \multirow{2}{*}{$2.67-5.32$}& \multirow{2}{*}{$3.71-8.10$} & \multirow{2}{*}{$1.59-3.74$}& \multirow{2}{*}{$2.29-6.79$}\\
& & & & \\
& NSM &$2.72-4.38$ &$3.30-5.90$ &$1.58-2.79$ & $2.12-4.44$\\
\hline
\multirow{3}{*}{$^{82}$Se} & \multirow{2}{*}{pnQRPA} & \multirow{2}{*}{$2.47-5.06$} & \multirow{2}{*}{$3.60-7.68$} & \multirow{2}{*}{$1.53-3.52$}& \multirow{2}{*}{$2.34-6.37$}\\
& & & & \\
& NSM &$2.36-3.72$ &$2.86-5.09$ &$1.34-2.46$ & $1.79-3.88$\\
\hline
\multirow{2}{*}{$^{96}$Zr} & \multirow{2}{*}{pnQRPA} & \multirow{2}{*}{$2.42-5.17$} & \multirow{2}{*}{$3.58-7.82$} & \multirow{2}{*}{$1.52-3.58$} & \multirow{2}{*}{$2.36-6.45$}\\
& & & & \\
\hline
\multirow{2}{*}{$^{100}$Mo} & \multirow{2}{*}{pnQRPA} & \multirow{2}{*}{$3.92-6.64$} & \multirow{2}{*}{$5.18-9.87$} & \multirow{2}{*}{$2.38-4.54$}& \multirow{2}{*}{$3.33-8.08$}\\
& & & & \\
\hline
\multirow{2}{*}{$^{116}$Cd} & \multirow{2}{*}{pnQRPA} & \multirow{2}{*}{$2.93-5.70$} & \multirow{2}{*}{$4.34-8.40$} & \multirow{2}{*}{$1.90-3.85$} & \multirow{2}{*}{$3.03-6.76$}\\
& & & & \\
\hline
\multirow{2}{*}{$^{128}$Te} & \multirow{2}{*}{pnQRPA} & \multirow{2}{*}{$2.00-4.89$} & \multirow{2}{*}{$3.09-7.56$} & \multirow{2}{*}{$1.31-3.38$} & \multirow{2}{*}{$2.06-6.26$}\\
& & & & \\
\hline
\multirow{3}{*}{$^{130}$Te} & \multirow{2}{*}{pnQRPA} & \multirow{2}{*}{$1.84-4.67$} & \multirow{2}{*}{$2.97-7.22$} & \multirow{2}{*}{$1.25-3.21$} & \multirow{2}{*}{$2.05-5.93$}\\
& & & & \\
& NSM &$1.62-3.96$ &$1.99-5.48$ &$0.92-2.57$ &$1.23-4.14$\\
\hline
\multirow{3}{*}{$^{136}$Xe} & \multirow{2}{*}{pnQRPA} & \multirow{2}{*}{$1.71-4.52$} & \multirow{2}{*}{$2.99-6.87$} & \multirow{2}{*}{$1.23-3.06$} & \multirow{2}{*}{$2.26-5.49$}\\
& & & & \\
& NSM &$1.29-2.90$ &$1.59-4.07$ &$0.72-1.90$ & $0.98-3.11$\\
\end{tabular}
\end{ruledtabular}
\label{tab:NMEs_from_correlations}
\end{table*}

\end{document}